\documentclass{elsart}

\usepackage{amsmath,amssymb}
\usepackage{graphicx}
\usepackage{cite}

\newcommand{\eVq}{\ensuremath{\text{eV}^2}}

\allowdisplaybreaks

\begin{document}


\begin{frontmatter}

\title{Status of Oscillation plus Decay of Atmospheric and
  Long-Baseline Neutrinos}

\author{M.~C.~Gonzalez-Garcia}
\address{%
  C.N.~Yang Institute for Theoretical Physics, SUNY at Stony Brook,
  Stony Brook, NY 11794-3840, USA}
\address{%
  Instituci\'o Catalana de Recerca i Estudis Avan\c{c}ats (ICREA),
  Departament d'Estructura i Constituents de la Mat\`eria, Universitat
  de Barcelona, 647 Diagonal, E-08028 Barcelona, Spain}

\author{Michele Maltoni}
\address{%
  Departamento de F\'isica Te\'orica \& Instituto de F\'isica
  Te\'orica UAM/CSIC, Facultad de Ciencias C-XI, Universidad
  Aut\'onoma de Madrid, Cantoblanco, E-28049 Madrid, Spain}

\begin{abstract}
    We study the interplay of neutrino oscillation and invisible decay
    in atmospheric and long-baseline neutrinos experiments. We perform
    a global analysis of the full atmospheric data from
    Super-Kamiokande together with long-baseline K2K and MINOS in
    these scenarios. We find that the admixture of $\nu_\mu \to
    \nu_\tau$ oscillations with parameters $\Delta m^2_{32} = 2.6
    \times 10^{-3}~\eVq$ and $\theta_{23} \sim 34^\circ$ plus decay of
    the heavy neutrino, $\nu_3$, with lifetime of the order
    $\tau_3/m_3 \sim 2.6 \times 10^{-12}$~s/eV provides a reasonable
    fit to atmospheric neutrinos, although this solution becomes more
    disfavored (dropping to the 99\% CL) once long-baseline data are
    included. Other than this local minimum, the analysis shows no
    evidence in favor of a non-vanishing neutrino decay width and an
    lower bound on the decay lifetime $\tau_3/m_3 \geq 9.3\times
    10^{-11}$~s/eV is set at 99\% CL. In the framework of Majoron
    models, this constraint can be translated into a bound on the
    Majoron coupling to $\nu_{3}$ and an unmixed very light sterile
    state, $|g_{s3}| \leq 8.6 \times 10^{-3}~(2.2~\text{eV}/{m_3})$.
\end{abstract}

\begin{keyword}
    neutrino oscillations \sep neutrino decay \sep Majoron models
    \PACS 14.60.Pq \sep 13.35.Hb \sep 14.80.Mz
\end{keyword}

\end{frontmatter}

Neutrino oscillations have entered an era in which the observations
from underground experiments obtained with neutrino beams provided to
us by Nature --~either from the Sun or from the interactions of cosmic
rays in the upper atmosphere~-- are confirmed and refined by
experiments using terrestrial beams from accelerators and nuclear
reactors~\cite{GonzalezGarcia:2007ib}.

In particular, with its high statistics data~\cite{Hosaka:2006zd}
Super-Kamiokande (SK) established beyond doubt that the observed
deficit in the $\mu$-like atmospheric events is due to $\nu_\mu \to
\nu_\tau$ oscillations, a result also supported by other atmospheric
experiments such as MACRO~\cite{Ambrosio:2001je} and
Soudan-2~\cite{Sanchez:2003rb}. This was further confirmed in
terrestrial experiments, first by the KEK to Kamioka long-baseline
(LBL) neutrino oscillation experiment (K2K)~\cite{Ahn:2006zza}, and
currently by the Fermilab to Soudan LBL experiment,
MINOS~\cite{Michael:2006rx}

Mass-induced neutrino oscillations are not the only possible mechanism
for $\nu_\mu \to \nu_\tau$ flavor transitions. They can also be
generated by a variety of nonstandard neutrino interactions or
properties~\cite{GonzalezGarcia:2007ib}.  Prior to the
highest-statistics SK data, some of these scenarios could provide a
good description --~alternative to $\Delta m^2$-induced
oscillations~-- of the atmospheric neutrino phenomenology.  In
particular, it was early noticed~\cite{Barger:1998xk} that a scenario
of very fast $\nu_j \to \nu_i$ oscillations plus invisible neutrino
decay $\nu_j \to \nu_i\, X$ could describe the $L/E$ dependence (where
$L$ is neutrino flight length and $E$ its energy) and the up-down
asymmetry of the contained events in SK if $\sin^2 \theta_{ij} \sim
0.87$ and $m_j / \tau_i \sim 1~\text{GeV} / D_E$ (where $D_E$ is the
diameter of the Earth). However, with more precise data, it was shown
that the description of the global contained event sample in this
scenario was worse than in the case of oscillations. Furthermore for
lifetimes favored by the contained event data very little $\nu_\mu$
conversion is expected for upgoing stopping muons in contradiction
with observation.  Based on these facts this mechanism was
subsequently ruled out in its simpler form~\cite{Lipari:1999vh,
Fogli:1999qt}.

The possibility of atmospheric neutrino decay was revisited in
Ref.~\cite{Barger:1999bg}, where the interplay of oscillations and
decay where discussed under the assumption that oscillations where
suppressed for atmospheric neutrinos, $\Delta m^2_{ij} <
10^{-4}~\eVq$, so that only the mixing plus decay effects were
relevant. It was found that a good fit to the contained and upgoing
muon atmospheric data at that time could be obtained for $\tau_i/m_i =
63~\text{km/GeV}$ and $\sin^2 \theta_{ij} = 0.30$.  Again, this
scenario became disfavored as the statistics accumulated by SK
increased. In particular, in Ref.~\cite{Ashie:2004mr} the SK
collaboration presented a study of the $\nu_\mu$ disappearance
probability as a function of $L/E$, finding evidence for a dip in the
$L/E$ distribution -- in agreement with the sinusoidal flavor
transition probability predicted by mass-induced oscillations. From
this they concluded that the mixing plus decay scenario of
Ref.~\cite{Barger:1999bg} provided a worse fit (by about $3.4\sigma$)
than the standard oscillation hypothesis to the observed event
distribution.

At present, besides the strong bounds from atmospheric data, the
observation of the $\nu_\mu$ energy spectrum both at K2K and MINOS
further constrains any $\nu_\mu$ flavor transition mechanism which
does not lead to the correct oscillatory behavior. 
However, this does not exclude that neutrino decay could play a role,
even if sub-dominant, in the atmospheric and LBL neutrino 
phenomenology, and in principle affect our determination of the 
neutrino parameters.  Conversely, if this is not the case, from a
joint analysis of oscillations and decay in atmospheric and LBL
experiments one can derive a robust bound on the neutrino decay
lifetime of the relevant states.  In this paper we address these
questions by performing a global analysis of the atmospheric and LBL
data with $\nu_\mu \to \nu_\tau$ transitions driven by neutrino masses
and mixing and allowing for neutrino decay. 

For the sake of concreteness we focus on scenarios with normal mass 
ordering of the neutrino states, $m_3 \geq m_2 \geq m_1$, in which the
heaviest neutrino ($\nu_3$ by convention) decays invisibly.  Since
$\nu_1$ and $\nu_2$ have large mixing with $\nu_e$, invisible decay of
these states is strongly constrained by the non-observation of its
effects in solar neutrinos~\cite{Joshipura:2002fb, Beacom:2002cb,
Bandyopadhyay:2002qg}, $\tau/m\gtrsim 10^{-4}$ s/eV, which makes it
completely unobservable in present atmospheric and LBL experiments.
Further simplification arises if one assumes that the decay products
are outside of the $(\nu_e,\, \nu_\mu,\, \nu_\tau)$ neutrino ensemble.
Indeed this is required in order for the decay to be fast enough
because in this case the mass difference for the decay
$\nu_3\rightarrow \nu_i X$, $\Delta m^2_{3i}$, may not be directly
constrained by oscillation data.  This is the case if, for example,
$\nu_3$ decays into a fourth much lighter sterile neutrino $\nu_s$
with which none of the active neutrinos mix~\cite{Barger:1999bg}.

With these assumptions, and neglecting the small allowed $\nu_e$
admixture in the oscillation~\cite{Apollonio:1999ae}, the atmospheric
and LBL neutrino evolution equation involves only two neutrino states 
$\vec\nu^T=(\nu_\mu,\nu_\tau)$.  $\nu_3$ decay can be accounted for by
introducing an imaginary part in the Hamiltonian which is proportional
to the only relevant decay width
\begin{equation}
    \label{eq:nudec}
    i\, \frac{d\vec\nu}{dx} = U_{23} \left[
    \frac{\Delta m_{32}^2}{4E}
    \begin{pmatrix}
	-1 & 0 \\	
	\hphantom{-}0 & ~1
    \end{pmatrix} 
    - i \frac{m_3}{2 \, \tau_3 \, E}
    \begin{pmatrix}
	0 & 0 \\
	0 & 1
    \end{pmatrix} \right] U_{23}^\dagger \, \vec\nu \,,
\end{equation}
where $\tau_3$ is the $\nu_3$
lifetime\footnote{Equation~\eqref{eq:nudec} also describes the
oscillation plus decay $\nu_3 \to \nu_i\, X$ even if $\nu_i$ has
admixtures of $\nu_\mu$ and $\nu_\tau$ provided that its energy is
degraded enough so that it does not contribute to the observed event
rates.} and $U_{23}$ is the rotation matrix of mixing angle
$\theta_{23}$.
Solving Eq.~\eqref{eq:nudec} one gets the survival probability of
$\nu_\mu$:
\begin{equation}
    \label{eq:pmumudec}
    P_{\mu\mu} = \cos^4 \theta_{23}
    + \sin^4 \theta_{23} e^{-\frac{m_3 \, L}{\tau_3 \, E}}
    + 2 \sin^2\theta_{23} \cos^2 \theta_{23} \,
    e^{-\frac{m_3\, L}{2 \tau_3\, E}}
    \cos\left( \frac{\Delta m_{32}^2 L}{2E} \right) \,.
\end{equation}
Equation~\eqref{eq:pmumudec} contains as limiting cases the scenarios
explored in Refs.~\cite{Barger:1998xk, Barger:1999bg} (up to a
relabeling $\nu_2 \leftrightarrow \nu_3$, or, equivalently,
$\sin\theta_{23} \leftrightarrow \cos\theta_{23}$).  If one assumes
that $\Delta m^2_{3i}\gg E/L$ the oscillating term in
Eq.~\eqref{eq:pmumudec} averages to zero and $P_{\mu\mu} = \cos^4
\theta_{23} + \sin^4 \theta_{23}\, e^{-\frac{m_3\, L}{\tau_3\, E}}$.
This was the decay model proposed in Ref.~\cite{Barger:1998xk}. In the
alternative scenario of Ref.~\cite{Barger:1999bg} $\nu_3$ decays into
a sterile state $\nu_j$ with which it does not mix.  In this case the
mass difference relevant for oscillations is unrelated to the mass
difference between the decaying and the daughter neutrino states so
one could have fast decays even if $\Delta m_{32}^2$ was very small.
In that limit $P_{\mu\mu} = \big( \cos^2 \theta_{23} + \sin^2
\theta_{23}\, e^{-\frac{m_3\, L}{2 \tau_3\, E}} \big)^2$. As mentioned
above both these limiting cases are now excluded.

\begin{figure}
    \includegraphics[width=0.98\linewidth]{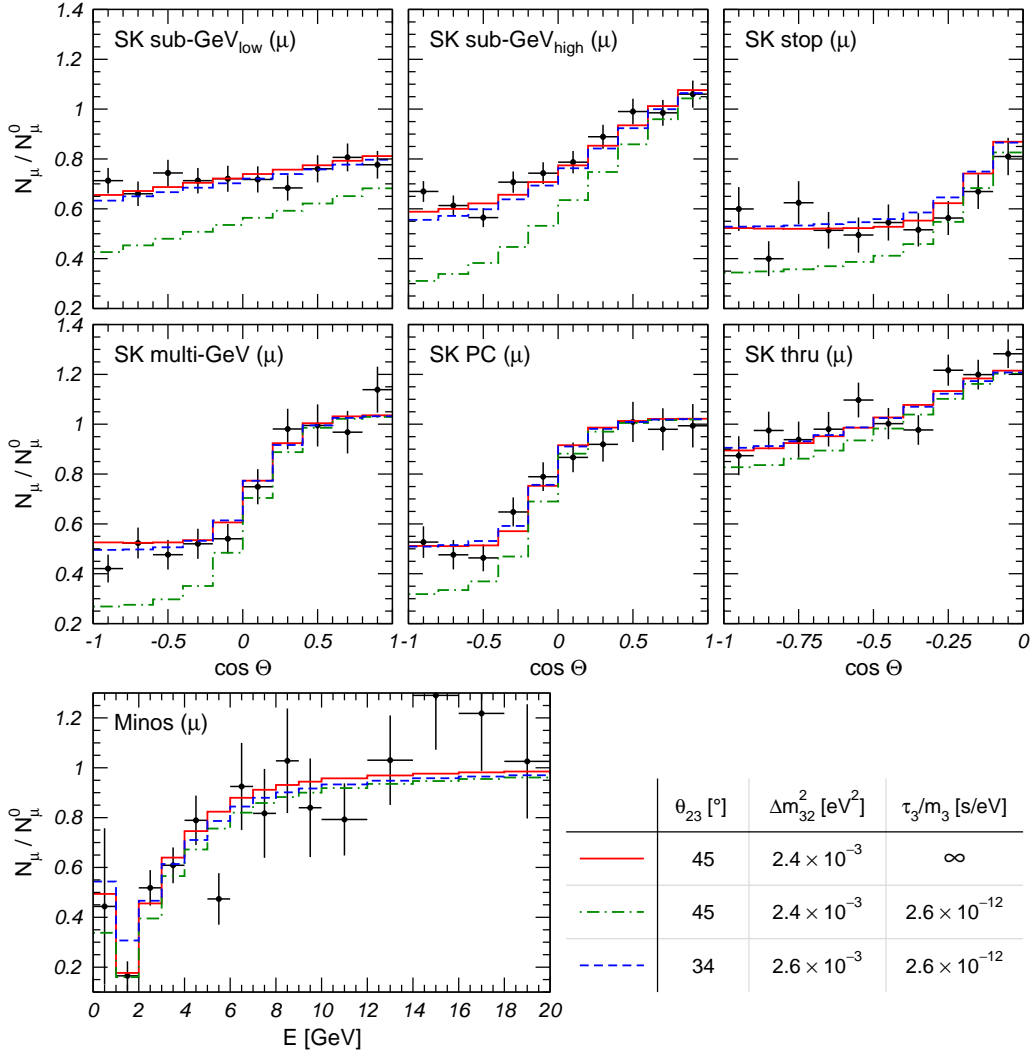}
    \caption{\label{fig:zenith}%
      Zenith-angle distributions for SK $\mu$-like events and energy
      spectrum at MINOS (normalized to the no-oscillation prediction) 
      for the global best fit (full line), for the local best fit with
      oscillations plus decay (dashed line), and for a solution with
      oscillation plus decay with maximal mixing (dash-dotted line).}
\end{figure}

In this work we have performed a global analysis of atmospheric and
long-baseline neutrino data in the general framework of oscillation
plus decay, as described by Eq.~\eqref{eq:pmumudec}, leaving free the
three parameters $\Delta m^2_{32}$, $\theta_{23}$ and $\tau_3/m_3$.
We have included all the SK-I and SK-II data as well as the latest K2K
and MINOS results. Concerning the analysis of atmospheric data, an
extensive description with all the technical details of our
calculations can be found in the Appendix of
Ref.~\cite{GonzalezGarcia:2007ib}.
As for MINOS, we convolve the unoscillated event spectrum given as a
function of the true neutrino energy (which we take from 
Refs.~\cite{minos1, minos2}) with the $P_{\mu\mu}$ survival
probability, and with a Gaussian smearing function to properly account
for the finite energy resolution of the detector. In this way we
calculate the charged-current event rates, which we add to the
neutral-current background (also taken from Ref.~\cite{minos2}) to
obtain the theoretical prediction for each energy bin. As can be seen
by comparing our event distribution for pure oscillations (lower panel
of Fig.~\ref{fig:zenith}) with the corresponding one from
MINOS~\cite{minos2}, our calculations show good agreement with the
MINOS Monte Carlo. The theoretical predictions are then fitted against
the experimental results, assuming a Poisson distribution with a total
systematic uncertainty of 4\%.

\begin{figure}
    \includegraphics[width=0.98\linewidth]{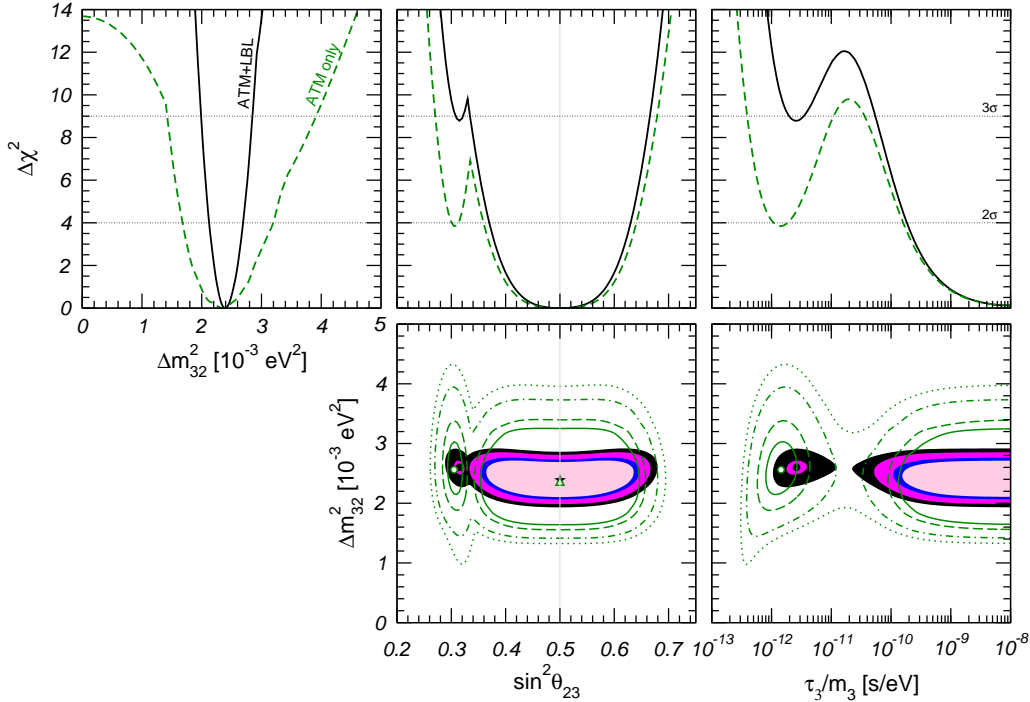}
    \caption{\label{fig:project}%
      Allowed regions from the analysis of atmospheric and LBL in
      presence of $\nu_\mu \to \nu_\tau$ oscillations plus $\nu_3$
      decay.  The three upper panels show the dependence of 
      $\Delta\chi^2$ on the parameters from the analysis of
      atmospheric only (dashed) and atmospheric+LBL (full line). The
      two lower panels show the two-dimensional projection of the
      allowed three-dimensional region after marginalization with
      respect to the undisplayed parameter. The different contours
      correspond to the two-dimensional allowed regions at 90\%, 95\%,
      99\% and $3\sigma$ CL. The lines (full regions) correspond to
      the atmospheric (atmospheric+LBL) analysis.}
\end{figure}

Our results are summarized Fig.~\ref{fig:project}, where we show
different projections of the allowed three-dimensional parameter space
after marginalization with respect to the undisplayed parameter. The
hollow regions in the two lower panels show the allowed domains at
90\%, 95\%, 99\% and $3\sigma$ CL from the analysis of the atmospheric
neutrino data alone; inclusion of the LBL experiments lead to the full
regions. The corresponding one-dimensional projections of the
$\Delta\chi^2$ functions are shown in the three upper panels.
From the figure we see that the best fit in this general oscillation
plus decay scenario corresponds to pure oscillations with
\begin{equation}
    \label{eq:bestfitglo}
    \Delta m^2_{32} = 2.4 \times 10^{-3}~\eVq \,, \qquad
    \theta = 45^\circ \,, \qquad
    \tau_3/m_3 \gg 10^{-8}~\text{s/eV} \,.
\end{equation}
However, the figure also shows that a reasonable fit to atmospheric
neutrino data is still possible in a oscillation plus decay scenario
with
\begin{equation}
    \label{eq:bestfitloc}
    \Delta m^2_{32} = 2.6 \times 10^{-3}~\eVq \,, \qquad
    \theta = 34^\circ \,, \qquad
    \tau_3/m_3 = 2.6 \times 10^{-12}~\text{s/eV} \,.
\end{equation}
This solution is well within the 90\% CL regions for the analysis of
atmospheric data alone (at $\Delta\chi^2_\text{ATM} = 3.8$ with
respect to the global best fit Eq.~\eqref{eq:bestfitglo}), although it
becomes more disfavored (at the 99\% CL level,
$\Delta\chi^2_\text{ATM+LBL} = 8.8$) when LBL data are also included
in the fit. The one-dimensional projection shows that the required
lifetime at this local best fit point lies at the boundary of the
$2\sigma$ ($3\sigma$) single parameter range allowed from atmospheric
(atmospheric+LBL) data. Indeed this solution is similar to that found
in Ref.~\cite{Barger:1999bg} for the decay and mixing but still
allowing for oscillations.  Our results show that forcing $\Delta
m^2_{32}\ll 10^{-3}$ rules out this solution well beyond $3\sigma$ (at
$\Delta\chi^2_\text{ATM} = 13.7$ with respect to the global best fit
Eq.~\eqref{eq:bestfitglo}), in agreement with the analysis of
SK~\cite{Ashie:2004mr}. LBL data definitively rules out the mixing
plus decay scenario with $\Delta\chi^2_\text{ATM+LBL} = 39$.

A better insight on the solutions in Eqs.~\eqref{eq:bestfitglo}
and~\eqref{eq:bestfitloc} can be obtained from Fig.~\ref{fig:zenith},
where we show the expected event distributions at SK and MINOS. As
seen in the figure both solutions yield rather similar results for the
atmospheric neutrino events. For the sake of comparison we also show
the corresponding distribution for a decay plus oscillation scenario
with $\Delta m^2$ and $\tau/m_3$ as in Eq.~\eqref{eq:bestfitloc} but
the mixing angle still maximal. This last curve illustrates how the
introduction of decay produces a strong deficit of atmospheric
$\nu_\mu$ events, which can be compensated by the deviation of the
mixing angle from maximal. Also, as easily seen from
Eq.~\eqref{eq:pmumudec} this compensation is only possible with a
mixing angle $\theta\leq 45$. The event distributions for MINOS also
show how the oscillation plus decay scenario cannot fully account for
the observed dip in the neutrino spectrum; this leads to the worsening
of the fit when including the LBL data. Thus this scenario will be
further tested by a more precise determination of the $\nu_\mu$
spectrum in MINOS.

Beyond this local minimum, the analysis shows no evidence in favor of
a non-vanishing neutrino decay width, thus it allows to set an upper
bound on the decay lifetime
\begin{equation}
    \label{eq:taulimit}
    \tau_3/m_3 \geq 2.9\; [0.93] \times 10^{-10}~\text{s/eV}
    \quad\Rightarrow\quad
    \tau_3 \geq 6.5\; [2.0] \times 10^{-10} \left(
    \frac{m_3}{2.2~\text{eV}} \right)\, \text{s}
\end{equation}
at the 90\% [99\%] CL where in the right hand side we have normalized
to the maximum allowed value on the absolute mass scale of the
neutrino from tritium $\beta$ decay experiments~\cite{Bonn:2001tw,
Lobashev:2001uu}, $m_{3} < 2.2~\text{eV}$.

We now turn to compare the bounds in Eq.~\eqref{eq:taulimit} with the
existing bounds from other experiments.  In order to make such
comparison, we must first specify the neutrino decay model.  The
reason for this is that most of the bounds in the literature are not
derived exclusively from effects due to the neutrino decay, but also
from effects associated with the presence of new neutrino interactions
which are responsible for its decay.
We will focus on the fast invisible Majorana neutrino decay $\nu_3 \to
\nu_s \, J$ induced by the interaction
\begin{equation}
    \label{eq:majdec}
    \mathcal{L}_I = i g_{ij} \, \bar{\nu}_{i} \gamma_5 \, \nu_{j}\, J \,,
\end{equation}
where $J$ is the Majoron (pseudoscalar) field~\cite{Gelmini:1980re,
Chikashige:1980ui, Acker:1992eh, Gelmini:1983ea, Choi:1991aa,
Joshipura:1992vn}, which has to be dominantly singlet, in order to
satisfy the constraints from the invisible decay width of
$Z$~\cite{GonzalezGarcia:1989zh}. Alternatively for Dirac neutrinos
one can have the decay channel $\nu_3 \to \bar \nu_s \chi$ induced by
a new neutrino interaction with a complex scalar field
$\chi$~\cite{Acker:1991ej}. In both cases the rest-frame lifetime of
$\nu_3$ for $m_3 \gg m_s$ is given by
\begin{equation}
    \tau_3 = \frac{16\pi}{g_{s3}^2 \, m_3} \,,
\end{equation}
where $g_{s3} = \cos\theta_{23}\, g_{s \tau } + \sin\theta_{23}\, g_{s
\mu}$ is the relation between the relevant coupling constants in the
mass and flavor basis (which for Majorana neutrinos is $g_{ij} =
U^T_{i\alpha}\, g_{\alpha\beta}\, U_{\beta j}$). For these modes, the
90\% [99\%] bounds on Eq.~\eqref{eq:taulimit} imply:
\begin{equation}
    \label{eq:glim}
    |g_{s3}| \leq 4.8\; [8.6] \times 10^{-3} \left(
    \frac{2.2~\text{eV}}{m_3} \right) \,.
\end{equation}
This bound can be directly compared with the constraints on the
$g_{\mu\alpha}$ and $g_{\tau\alpha}$ (for any flavor or sterile state
$\alpha$) couplings that have been derived from their effect in meson
and charged lepton decay.  The most updated
analysis~\cite{Lessa:2007up} yields the model independent 90\% bounds
\begin{equation}
    |g_{\mu\alpha}| \leq 9.4 \times 10^{-3}\,, \qquad
    |g_{\tau\alpha}| \leq 0.33\,.
\end{equation}
Also, limits from decay and scattering of Majorons inside supernova
yield bounds on $|g_{\alpha\beta}|$ because for large couplings the
supernova energy is drained due to Majoron emission and no explosion
occurs. However, for very large coupling the Majoron becomes trapped
inside the supernova and no constraint is
possible~\cite{Kachelriess:2000qc, Tomas:2001dh}. As a consequence
both ranges
\begin{equation}
    \label{eq:snbound}
    |g_{\alpha\beta}| < 3 \times 10^{-7}
    \qquad\text{or}\qquad
    |g_{\alpha\beta}| > 2 \times 10^{-5}
\end{equation}
are allowed~\cite{Kachelriess:2000qc}.

In Ref.~\cite{Hannestad:2005ex} a very strong limit was derived from
the requirement that the neutrinos are free-streaming at the time of
the photon decoupling, as deduced by precise measurements of the CMB
acoustic peaks, $|g_{ij}| \lesssim 0.61 \times 10^{-11}~(50~\text{meV}
/ m_i)^2$. However the robustness of this conclusion has been
questioned in~\cite{Bell:2005dr}.

Future experiments can improve the bounds on $g_{s \mu}$ and $g_{s
\tau}$ by orders of magnitude, in particular from their cosmological
effects~\cite{Serpico:2007pt} and from the observation of diffuse
supernova neutrino background~\cite{Ando:2003ie, Fogli:2004gy}. Till
then the bounds derived in this work, Eqs.~\eqref{eq:taulimit}
and~\eqref{eq:glim}, from the analysis of atmospheric and LBL neutrino
data are the strongest applicable to the $\nu_3$ neutrino state with
masses $\mathcal{O}(\text{eV})$. In the flavor basis they represent
the strongest bounds on the Majoron coupling to $\nu_\tau$--$\nu_s$ in
the full range $2.2\leq m_3\leq 0.05$ eV allowed for the normal
ordering of the neutrino mass states $m_3 \geq m_2 \geq m_1$ with
$m_3\gg m_s$.

\bigskip

We thank Y.~Nir for careful reading of the manuscript and comments. 
This work is supported by National Science Foundation grant
PHY-0354776 and by Spanish Grants FPA-2004-00996, FPA2006-01105 and
FPA-2007-66665-C02-01.  MM is supported by MCI through the Ram\'on y
Cajal program, by the Comunidad Aut\'onoma de Madrid through the
HEPHACOS project P-ESP-00346, and by the European Union through the
ENTApP network of the ILIAS project RII3-CT-2004-506222.


\end{document}